\documentstyle[12pt,aaspp4]{article}

\begin{document}

\title{RELATIVISTIC DYNAMOS IN MAGNETOSPHERES \\
 OF ROTATING COMPACT OBJECTS}  
\author{AKIRA TOMIMATSU}  
\affil{Department of Physics, Nagoya University, Chikusa-ku, Nagoya 464-8602, Japan \\
E-mail: atomi@allegro.phys.nagoya-u.ac.jp}

\begin{abstract}
The kinematic evolution of axisymmetric magnetic fields in rotating magnetospheres of relativistic compact objects is analytically studied, based on relativistic Ohm's law in stationary axisymmetric geometry. By neglecting the poloidal flows of plasma in simplified magnetospheric models, we discuss self-excited dynamos due to the frame-dragging effect (originally pointed out by Khanna \& Camenzind), and we propose alternative processes to generate axisymmetric magnetic fields against ohmic dissipation. The first process (which may be called induced excitation) is caused by the help of a background uniform magnetic field in addition to the dragging of inertial frames. It is shown that excited multipolar components of poloidal and azimuthal fields are sustained as stationary modes, and outgoing Poynting flux converges toward the rotation axis. The second one is self-excited dynamo through azimuthal convection current, which is found to be effective if plasma rotation becomes highly relativistic with a sharp gradient in the angular velocity. In this case no frame-dragging effect is needed, and the coupling between charge separation and plasma rotation becomes important. We discuss briefly the results in relation to active phenomena in the relativistic magnetospheres. 
\end{abstract}


DPNU-99-26

\newpage
\section{INTRODUCTION}

Interesting high-energetic phenomena have been observed in various compact astrophysical systems, such as pulsars, X-ray binaries and active galactic nuclei. Though the energy-release processes via radiation emission and jet formation have not been fully understood as yet, strong magnetic fields near the central objects can be a crucial component for explaining the observational features, and many works have been devoted to the developments of relativistic magnetohydrodynamical (MHD) models. In particular, Khanna \& Camenzind (1994\markcite{KC94}, 1996a\markcite{KC96a}) have recently proposed a self-excitation mechanism of axisymmetric magnetic fields, based on relativistic Ohm's law in Kerr geometry. This effect which is called self-excited gravitomagnetic dynamo is due to the coupling between relativistic frame-dragging of a rotating central object and rotational motion of surrounding plasma, and such dynamo action is expected to play an important role in the astrophysical phenomena as a trigger of relativistic plasma flows. Unfortunately, numerical calculations done by Brandenburg\markcite{Br96} (1996) have shown that in a wide set of standard thin disk models around a rotating black hole no magnetic field can be maintained against ohmic dissipation. Cowling's antidynamo theorem for axisymmetric magnetic fields still holds even near a rotating black hole in the situations previously considered. Therefore, a new viewpoint will be necessary if the kinematic theory of resistive MHD presented by Khanna \& Camenzind (1994, 1996a) is applied to the problem of generation of axisymmetric magnetic fields (see also N\'{u}\~{n}ez\markcite{Nu97} 1997).

In this paper we do not adhere to the accretion disk models, but we pursue analytically the processes permissible in rotating magnetospheres of compact objects. Though in the kinematic treatment of neglecting the feedback of magnetic fields due to the Lorentz force on plasma velocity the basic field equations become much simpler in comparison with the full MHD theory, some additional approximations are required to make any analytical approach possible. We would like to restrict attention to the simplified cases in which only the essential interactions between poloidal and azimuthal magnetic fields for allowing the existence of growing or stationary modes are preserved.

The plasma injected into the rotating magnetosphere will partially accrete onto the central object and partially escape to infinity. The poloidal velocity of plasma flows, however, can remain sub-Alfvenic in some intermediate (quasi-equilibrium) region between the outer light cylinder and the surface of the central object, where the plasma angular velocity may be different from the Keplerian one, because gravitational forces do not dominate in comparison with any other interactions (see, e.g., Camenzind\markcite{Ca87} 1987; Takahashi et al.\markcite{Ta90} 1990). The stationary axisymmetric structure of the magnetosphere is mainly determined in the framework of ideal MHD theory under the frozen-in condition. We consider the evolution of electromagnetic fields perturbed by the presence of small magnetic diffusivity. Though the motion of the plasma is assumed to be non-perturbed, a complicated dynamical evolution of electromagetic fields due to the poloidal flows will occur. Therefore, by fixing the poloidal velocity to be zero in the quasi-equilibrium region, we study the generation of magnetic fields which goes on slowly in a long diffusion timescale. Our purpose is to point out some basic aspects of the perturbed fields governed by the plasma rotation and the frame-dragging effect. We obtain the main results such that (i) if a background uniform magnetic field exists, it can sustain excited poloidal and azimuthal multipolar modes even in slow-rotation cases, and (ii) a sufficient charge separation generated by plasma rotation with a relativistic speed can cause self-excited dynamo without any frame-dragging effect. We discuss these processes of generation of magnetic fields (which have been missed in the above-mentioned numerical models) in relation to active phenomena observed in the relativistic magnetospheres. 

In the following  we use units such that $c=G=1$, and the axisymmetric stationary metric denoted by $g_{ab}$ has signs ($-$ + + +).

\section{KINEMATIC EQUATIONS FOR AXISYMMETRIC DYNAMOS}

A contribution of accreting plasma and electromagnetic fields around a rotating compact object to the stationary axisymmetric gravitational field remains negligibly small. Therefore, we can always study the MHD interaction under a fixed gravitational field with the line element of the form 
\begin{equation}
ds^{2} \ = \ g_{tt}dt^{2}+2g_{t\phi}dtd\phi+g_{\phi\phi}d\phi^{2}+g_{rr}dr^{2}+g_{\theta\theta}d\theta^{2} \ ,
\end{equation}
where $r$, $\theta$ and $\phi$ are the spherical coordinates, and the metric $g_{ab}$ is assumed to be independent of the time coordinate $t$ and the azimuthal angle coordinate $\phi$. The angular velocity of the dragging of inertial frames is denoted by $\omega\equiv -g_{t\phi}/g_{\phi\phi}$. We define $g$ to be the determinant of $(g_{ab})$, and for the lapse function $\alpha\equiv\sqrt{-g_{tt}+(g_{t\phi}^{2}/g_{\phi\phi})}$ we have $\sqrt{-g}=\alpha\sqrt{g_{\phi\phi}g_{rr}g_{\theta\theta}}$. We do not limit the metric to the Kerr form for later discussions. Further, we do not use explicitly the 3+1 formalism developed by Thorne, Price, \& Macdonald\markcite{Th86} (1986), and we treat relativistic Ohm's law in a covariant form written by
\begin{equation}
F_{ab}u^{b} \ = \ 4\pi\eta(j_{a}-Qu_{a}) \ ,   
\end{equation} 
where $\eta$ is the magnetic diffusivity. According to the kinematic MHD theory the plasma 4-velocity $u^{a}(r,\theta)$ is also fixed, and $Q\equiv-j^{a}u_{a}$ is the electric charge density measured by an observer comoving with the plasma. The electric current density $j^{a}$ should be rewritten by the electromagnetic field $F_{ab}$ via the Maxwell equations $4\pi j^{a}=\nabla_{b}F^{ab}$, where $\nabla_{b}$ denotes the covariant derivative with respect to the metric $g_{ab}$. 

Because we consider time-dependent fields under the assumption of axisymmetry, the poloidal magnetic components $F_{r\phi}$ and $F_{\theta\phi}$ and the azimuthal electric component $F_{t\phi}$ are given by the single scalar potential $\Psi(t,r,\theta)$ via the equation
\begin{equation}
F_{a\phi} \ = \ \partial_{a}\Psi \ . 
\end{equation}
Then the four field variables $\Psi$, $F^{rt}$, $F^{\theta t}$ and $F^{r\theta}$ remain to be solved, and the equation added to relativistic Ohm's law is the azimuthal part of the Faraday law 
\begin{equation}
\partial_{t}F_{r\theta}+\partial_{r}F_{\theta t}+\partial_{\theta}F_{tr} \ = \ 0 \ . 
\end{equation}

These field equations for the kinematic evolution are still too complicated to discuss analytically the behaviors of solutions. Hence, the following investigation is limited to the models with no poloidal flow, i.e.,
\begin{equation}
u^{r} \ = \ u^{\theta} \ = 0 \ ,
\end{equation} 
which will be justified if plasma is located in a quasi-equilibrium region slightly distant from the surface of the central object, and a dynamical evolution of electromagnetic fields caused by the poloidal plasma flows with sub-Alfvenic velocities is not essential to the problem of dynamo action. Further, we assume $\eta$ to be a very small constant, and the time variation of fields is described by a long diffusion timescale such that $t\sim r^{2}/\eta$. Then it is convenient to introduce the variable $T\equiv\eta t$ instead of $t$. As a result of this assumption concerning the order of $\eta$ the field equations can have consistent solutions, if the ratio of the amplitudes is understood to be 
\begin{equation}
\eta F_{r\theta}/\Psi \ = O(1) \ . 
\end{equation}

Now let us give the field equations which are simplified according to the above-mentioned approximations. By virtue of equation (5) the poloidal part of equation (2) leads to  
\begin{equation}
F^{At}u_{t}+F^{A\phi}u_{\phi} \ = \ \eta j^{A} \ , 
\end{equation}
where the poloidal current density is approximately given by 
\begin{equation}
\eta j^{A} \ = \ \frac{1}{\sqrt{-g}}\epsilon^{AB}\partial_{B}F \ , \ \ \epsilon^{r\theta} \ = \ -\epsilon^{\theta r} \ = \ 1 \ , 
\end{equation}
in which the displacement current is neglected. (Hereafter the superscripts or subscripts written by $A$ and $B$ mean the coordinates $r$ and $\theta$.)  Because we have $F_{A\phi}=\partial_{A}\Psi$, these relations are used to express $F^{At}$ and $F^{A\phi}$ by $\Psi$ and $F\equiv\eta\sqrt{-g}F^{r\theta}$, for example, in the approximated form of the proper charge density $Q$ given by  
\begin{equation}
4\pi Q \ = \ -\frac{u_{a}}{\sqrt{-g}}\partial_{b}(\sqrt{-g}F^{ab}) \ \simeq \ F^{tA}\partial_{A}u_{t}+F^{\phi A}\partial_{A}u_{\phi} \ ,  
\end{equation}
which should be derived from equation (2) by using the current conservation $\nabla_{a}j^{a}=0$ and the inequality $|Q|\gg|\eta\nabla_{a}(Qu^{a})|$. Note that for the stationary and axisymmetric velocity field $u_{a}$ we have $\partial_{t}u_{a}=\partial_{\phi}u_{a}=0$, while the rotational motion can generate the non-vanishing components $\partial_{A}u_{t}$ and $\partial_{A}u_{\phi}$ for $A=r, \theta$. Then, except in the case that both $u_{t}$ and $u_{\phi}$ are constant, $\partial_{b}u_{a}$ becomes non-symmetric under the permutation of $a$ and $b$ to assure the validity of equation (9) for the estimation of charge separation.

From the Maxwell equations we also obtain approximately the azimuthal current density $j_{\phi}$, in order to substitute it into the azimuthal part of equation (2) of the form 
\begin{equation}
u^{t}\partial_{T}\Psi \ = \ -4\pi(j_{\phi}-Qu_{\phi}) \ , 
\end{equation}
with equation (9) for the proper charge density $Q$. Then, we arrive at the final result of the evolution equation for $\Psi$ 
\begin{equation}
\partial_{T}\Psi \ = \ S_{1}+S_{2}+S_{3} \ , 
\end{equation}
where 
\begin{equation}
S_{1} \ = \ \frac{g_{\phi\phi}}{u^{t}\sqrt{-g}}\partial_{A}(\frac{\sqrt{-g}}{g_{\phi\phi}}\partial^{A}\Psi) \ , 
\end{equation}
\begin{equation}
S_{2} \ = \ \frac{u_{\phi}}{(\alpha u^{t})^{2}g_{\phi\phi}}\partial^{A}\Psi(g_{\phi\phi}\partial_{A}\omega+u_{t}\partial_{A}u_{\phi}-u_{\phi}\partial_{A}u_{t}) \ , 
\end{equation}
\begin{equation}
S_{3} \ = \ \frac{1}{(\alpha u^{t})^{2}\sqrt{-g}}\epsilon^{AB}\partial_{A}F
\{g_{\phi\phi}\partial_{B}\omega-u_{\phi}(\partial_{B}u_{t}+\omega\partial_{B}u_{\phi})\} .  
\end{equation}
The final term $S_{3}$ can contribute to the excitation of $\Psi$ through the coupling to $F$. The ohmic diffusion is mainly due to the first term $S_{1}$, and the role of $S_{2}$ (i.e., self-generation or self-destruction) will depend on the topology of $\Psi$.  

The Faraday law (4) is the evolution equation for $F$, which approximately reduces to the form
\begin{equation}
\partial_{T}F \ = \ \frac{\alpha^{2}g_{\phi\phi}}{\sqrt{-g}}\{\partial_{A}(\frac{\sqrt{-g}}{\alpha^{2}u^{t}g_{\phi\phi}}\partial^{A}F)
-\epsilon^{AB}\partial_{A}\Psi\partial_{B}\Omega\} \ , 
\end{equation}
where $\Omega\equiv u^{\phi}/u^{t}$ is the plasma angular velocity. Note that the right-hand side of equation (15) is also decomposed into the terms for diffusion and amplification of $F$. In the following sections we will study the coupled equations (11) and (15) for $\Psi$ and $F$ to see the efficiency of excitation mechanisms.  

\section{THE FRAME-DRAGGING EFFECT}

In the case of no poloidal velocity $u^{A}=0$ we can give the specific energy and angular momentum of plasma denoted by $-u_{t}$ and $u_{\phi}$ as follows, 
\begin{equation}
-u_{t} \ = \ \gamma\{\alpha^{2}+g_{\phi\phi}\omega(\Omega-\omega)\} \ , \ \ 
u_{\phi} \ = \ \gamma g_{\phi\phi}(\Omega-\omega) 
\end{equation}
where $\gamma\equiv u^{t}=1/\sqrt{\alpha^{2}-g_{\phi\phi}(\Omega-\omega)^{2}}$ is the Lorentz factor of rotating plasma. If the plasma is co-rotating with the angular velocity of the background magnetosphere, the Lorentz factor $\gamma$ becomes very large in a region close to the light cylinder surface, and the term originated from the convection current density $Qu_{\phi}$ will dominate in $S_{3}$. In this section we would like to restrict attention to the frame-dragging effect, by neglecting any contribution of such charge separation under the condition $\Omega=\omega$. 

\subsection{The $\omega$-$\Omega$ Dynamo}

Recall that in the numerical models of $u_{\phi}=0$ calculated by Brandenburg (1996) the dynamo action can work only if $\omega$ is taken to be artificially large. To see roughly this result from equations (11) and (15) with $u_{\phi}=0$, let us consider simplified forms of the metric components as functions of the coordinate $\theta$ such that $g_{\phi\phi}$ and $-g$ are proportional to $\sin^{2}\theta$, and $\omega$, $g_{rr}$ and $g_{\theta\theta}$ are independent of $\theta$. (Now the Lorentz factor $u^{t}=1/\alpha$ depends only on $r$. This simplified metric may be regarded as the Kerr metric in slow-rotation approximation.) Then, by setting the time behavior of $\Psi$ and $F$ to be $\exp(\mu T)$, we obtain
\begin{equation}
\frac{g_{\theta\theta}}{1-x^{2}}(\frac{\mu}{\alpha}-L_{1})\Psi \ = \ \partial_{x}^{2}\Psi-\frac{\sigma}{\alpha}\partial_{x}F \ , 
\end{equation}
and
\begin{equation}
\frac{g_{\theta\theta}}{1-x^{2}}(\frac{\mu}{\alpha}-L_{2})F \ = \ \partial_{x}^{2}F+\alpha\sigma\partial_{x}\Psi \ , 
\end{equation}
where $x\equiv\cos\theta$. The efficiency of dynamo action will be determined by the behavior of $\sigma(r)$ dependent on the metric components as follows,
\begin{equation}
\sigma \ = \ -\frac{g_{\phi\phi}g_{\theta\theta}}{\sin\theta\sqrt{-g}}\frac{d\omega}{dr} \ . 
\end{equation}
Further, $L_{1}\Psi$ and $L_{2}F$ in the left-hand sides of equations (17) and (18) represent the ohmic diffusion of $\Psi$ and $F$ in radial direction, which are given by 
\begin{equation}
L_{1}\Psi \ = \ \frac{g_{\phi\phi}}{\sqrt{-g}}\partial_{r}(\frac{\sqrt{-g}}{g_{\phi\phi}g_{rr}}\partial_{r}\Psi) \ , 
\end{equation}
and
\begin{equation}
L_{2}F \ = \ \frac{\alpha g_{\phi\phi}}{\sqrt{-g}}\partial_{r}(\frac{\sqrt{-g}}{\alpha g_{\phi\phi}g_{rr}}\partial_{r}F) \ . 
\end{equation}

We can decompose $\Psi$ and $F$ into modes symmetric or antisymmetric with respect to the equatorial plane (N\'{u}\~{n}ez\markcite{Nu96} 1996), and in this paper we would like to consider only the configuration of magnetic fields with the symmetry such that $\Psi(-x)=\Psi(x)$ and $F(-x)=-F(x)$. (This corresponds to dipole-type topology of $\Psi$. Quadrupole-type fields may be more easily excited near the equatorial plane. Here we do not pursue this possibility.) For physical modes the functions $\Psi$ and $F$ should satisfy the boundary conditon that both $\Psi/(1-x^{2})$ and $F/(1-x^{2})$ remain finite on the polar axis ($x^{2}=1$). Then, if the diffusion terms $L_{1}\Psi$ and $L_{2}F$ are neglected, it is easy to obtain 
\begin{equation}
\Psi \ = \ \Psi_{0}\{1-(-1)^{n}\cos(\sigma x)\} \ , \ \ F \ = \ (-1)^{n}\alpha\Psi_{0}\sin(\sigma x)   
\end{equation}
as a stationary eigenmode with $\mu=0$. By virtue of the boundary condition for $\Psi$ and $F$ the eigenvalue of $\sigma$ is given by $\sigma=n\pi$, where $n=1, 2, \cdots$. Of course, if one takes account of a diffusion effect due to the terms $L_{1}\Psi$ and $L_{2}F$, the minimum value of $\sigma$ should become larger than $\pi$. However, for the Kerr metric on the equatorial plane, we can estimate the value of $\sigma$ to be 
\begin{equation}
\sigma \ = \ \frac{2Ma(3r^{2}+a^{2})}{r^{4}+a^{2}(r^{2}+2Mr)} \ \leq \ 2 \ , 
\end{equation}
where $M$ and $a$ are the mass and rotation parameters, respectively. Therefore, one can expect no self-excitation of fields to occur near the Kerr black hole. In this sense the rotation of the black hole turns out to be too slow to excite the dynamo action. 
\subsection{Induced Excitation}

Now let us propose an alternative process which can work even in the slow-rotation case and may be called {\it induced excitation} instead of {\it self-excitation}. The key assumption is the existence of a background poloidal field denoted by $\Psi_{B}$ as a stationary solution of the vacuum Maxwell equations $\nabla_{b}F^{ab}=0$. (The typical example is given by Wald's\markcite{Wa74} (1974) solution for the Kerr hole immersed in a uniform magnetic field with the form $\Psi_{B}=B_{0}\{ag_{t\phi}+(g_{\phi\phi}/2)\}$.) If plasma is injected into the magnetosphere, the structure should be deformed by the motion of plasma. However, the original vacuum field in the background magnetosphere can remain dissipation-free and play a role of a stationary source field in equations (17) and (18). Then, we will be able to find a stationary solution written by $\Psi=\Psi_{B}+\Psi_{L}$ and $F=F_{L}$. These perturbed parts $\Psi_{L}$ and $F_{L}$ represent localized fields with amplitudes decreasing for large $r$, and the dynamical balance between the ohmic dissipation and the excitation via the frame-dragging effect for $\Psi_{L}$ and $F_{L}$ is induced by the background field $\Psi_{B}$. (This process has been also mentioned in Khanna \& Camenzind\markcite{KC96b} (1996b) and Khanna\markcite{Kh97} (1997), and the numerical examples have been presented by Khanna\markcite{Kh98c} (1998c).) 

To verify this induced excitation as a viable process, we write the stationary fields $\Psi$ and $F$ satisfying equations (17) and (18) in the expansion forms 
\begin{equation}
\Psi \ = \ (1-x^{2})\sum_{n=0}^{\infty}q_{2n}(r)\frac{dP_{2n+1}(x)}{dx} \ , 
\end{equation}
and
\begin{equation}
F \ = \ (1-x^{2})\sum_{n=0}^{\infty}q_{2n+1}(r)\frac{dP_{2n+2}(x)}{dx} \ , 
\end{equation}
according to the boundary condition on the polar axis and the symmetry with respect to the equatorial plane. By the help of the recurrence relations for Legendre polynomials $P_{n}$ we have the equations for the coefficients $q_{n}$ as follows, 
\begin{equation}
g_{\theta\theta}(L_{1}q_{2n})-(2n+1)(2n+2)q_{2n} \ = \ \frac{\sigma}{\alpha}(c_{2n+1}q_{2n+1}-c_{2n-1}q_{2n-1}) \ , 
\end{equation}
and
\begin{equation}
g_{\theta\theta}(L_{2}q_{2n+1})-(2n+2)(2n+3)q_{2n+1} \ = \ -\alpha\sigma(c_{2n+2}q_{2n+2}-c_{2n}q_{2n}) \ ,  
\end{equation}
where $c_{n}=(n+1)(n+2)/(2n+3)$. It is clear that the higher multipolar modes are generated from the lower ones through the action of the dragging of inertial frames denoted by $\sigma$. 
 
Our main purpose is to point out a remarkable difference of the efficiency between the self-excitation and the induced one. Hence, for further analytical study, we consider a distant region where the metric components are approximately given by
\begin{equation}
\alpha \ = \ 1 \ , \ \ g_{\phi\phi} \ = \ r^{2}\sin^{2}\theta \ , \ \ g_{rr} \ = \ g_{\theta\theta}/r^{2} \ = \ 1 \ , 
\end{equation}
keeping the dragging of inertial frames written as 
\begin{equation}
\omega \ = \ 2J/r^{3}
\end{equation}
for the angular momentum $J$ of a central object. Equation (29) leads to 
\begin{equation}
\sigma \ = \ 6J/r^{2} \ \ll \ 1 \ . 
\end{equation}
In order to assure $\sigma=-r^{2}d\omega/dr$ to remain very small even for a small $r$ in the following calculation, one may assume the behavior
\begin{equation}
\sigma \ = \ (r/r_{c})^{2}\sigma_{c} \ , \ \ \sigma_{c} \ = \ 6J/r_{c}^{2}  
\end{equation} 
in the inner region $r<r_{c}$, where $r_{c}$ will be of the order of the radius of a central object. 

In the slow-rotation limit the recurrence relations (26) and (27) for $n\geq1$ reduce to the approximated form
\begin{equation}
\frac{d^{2}q_{n+1}}{dr^{2}}-\frac{(n+2)(n+3)}{r^{2}}q_{n+1} \ = \ (-1)^{n}\frac{\sigma c_{n}}{r^{2}}q_{n} \ , 
\end{equation}
because we can neglect $q_{n+2}$ in comparison with $q_{n}$ in the right-hand sides. (Consistently with equation (32), the ratio $q_{n+1}/q_{n}$ is assumed to be of the order of $\sigma$, and the convergence of the expansion forms equations (24) and (25) is assured.)  Now, if $q_{n}$ is known, it is easy to obtain the higher multipolar mode $q_{n+1}$. Note that in the flat spacetime the azimuthal component of magnetic field measured in orthonormal frame is given by $B_{T}=F/\eta r\sin\theta$. Then, $q_{2n+1}/r$ should vanish in the limit $r\rightarrow\infty$ and be regular in the limit $r\rightarrow0$. The function $q_{2n}$ for any localized poloidal flux should satisfy the same boundary conditions. For $\sigma=0$ we have the two independent solutions for each $q_{n+1}$ as follows,
\begin{equation}
q_{n+1} \ = \ r^{-(n+2)} \ , \ \ \ q_{n+1} \ = \ r^{n+3} \ , 
\end{equation} 
violating either the outer boundary condition or the inner one. If there exists a localized field corresponding to a lower multipolar mode $q_{n}$, however, the frame-dragging effect giving $\sigma\neq0$ can generate the higher one  
\begin{equation}
q_{n+1} \ = \ r^{-(n+2)}\int_{0}^{r}b_{n}(\rho)\rho^{2n+4}d\rho \ , 
\end{equation}
where
\begin{equation}
b_{n}(r) \ = \ \int_{r}^{\infty}(-1)^{n+1}\frac{\sigma c_{n}}{\rho^{n+4}}q_{n}(\rho)d\rho \ . 
\end{equation}
Therefore, the key problem is the generation of the lowest mode $q_{0}$, for which we obtain the equation
\begin{equation}
\frac{d^{2}q_{0}}{dr^{2}}-\frac{2}{r^{2}}q_{0} \ = \ \frac{6\sigma}{5r^{2}}q_{1} \ ,
\end{equation}
because $c_{n}=0$ for $n=-1$. Note that the remaining source field for $q_{0}$ is only the lowest mode $q_{1}$ of the azimuthal magnetic field, satisfying the equation
\begin{equation}
\frac{d^{2}q_{1}}{dr^{2}}-\frac{6}{r^{2}}q_{1} \ = \ \frac{2\sigma}{3r^{2}}q_{0} \ . 
\end{equation}
Both modes should be self-consistently generated according to these coupled equations. 

As was previously mentioned, a localized solution as a result of self-excited dynamo will be prohibited for a small $\sigma$. For example, we can check the suppression of the action even in the extreme case of assuming a sharp gradient in the angular velocity $\omega$ at $r=r_{c}$: The value of $\omega$ is given by a nonzero constant $\Delta\omega$ in the inner region $r<r_{c}$, while it becomes zero in the outer region $r>r_{c}$. In this case, though the localized modes $q_{0}$ and $q_{1}$ can be continuous with values denoted by $q_{c0}$ and $q_{c1}$, the gradients $dq_{0}/dr$ and $dq_{1}/dr$ must have discontinuous gaps estimated to be $-3q_{c0}/r_{c}$ and $-5q_{c1}/r_{c}$, respectively. (We have $q_{n}\sim r^{n+2}$ for $r<r_{c}$, while $q_{n}\sim r^{-(n+1)}$ for $r>r_{c}$.) Then, integrating equations (36) and (37) over the narrow region with the sharp gradients, we obtain the eigenvalue of the discontinuous gap given by $r_{c}\Delta\omega=5\sqrt{5}/2$. 

N\'{u}\~{n}ez (1997) has also treated a case with a sharp gradient in $\Omega$ by using equation (29) for $\omega$, and he has claimed the existence of growing modes for a smaller discontinuous gap $r_{c}\Delta\Omega\leq1$. This difference will be mainly due to the estimation of the diffusion term given by $\partial^{2}\Psi/\partial r^{2}$, which was rewritten into the form $\epsilon^{2}\partial^{2}\Psi/\partial x^{2}$ under the scale transformation $r=r_{c}-(\epsilon/2)+\epsilon x$. It seems to be unacceptable that the amplitude of the diffusion term is suppressed by the small factor $\epsilon^{2}$. In fact, the straightforward application of the scale transformation will lead to $\partial^{2}\Psi/\partial r^{2}=\epsilon^{-2}\partial^{2}\Psi/\partial x^{2}$. In our calculation the diffusion terms $d^{2}q_{0}/dr^{2}$ and $d^{2}q_{1}/dr^{2}$ are estimated to be very large by virtue of the steep change of the gradients $dq_{0}/dr$ and $dq_{1}/dr$. Then, the ohmic dissipation due to the diffusion terms dominate in the evolution, unless the discontinuous gap $\Delta\omega$ itself becomes unphysically large. 
 
Therefore, let us consider the induced excitation for the physically plausible behavior of $\sigma$ given by equations (30) and (31). In the case $\sigma=0$, the two independent solutions for $q_{0}$ have the forms $r^{2}$ and $r^{-1}$ corresponding to a uniform magnetic field and a dipole one, which will be relevant to the magnetospheres of black holes and neutron stars. Of course, these vacuum fields should be regarded as background fields in the magnetospheres, of which the origin cannot be attributed to the dynamo action discussed here. Our problem is rather to study the field generation against the ohmic dissipation in the background magnetospheres. We limit the analysis to the case of the uniform background field with the strength $B_{0}$. Then, for $\sigma\neq0$ due to the frame-dragging effect the solution $q_{0}$ is modified into the form
\begin{equation}
q_{0} \ = \  \frac{B_{0}r^{2}}{2}+\sigma_{c}^{2}p_{0} \ .
\end{equation}
To obtain $q_{1}$ in the first order of $\sigma_{c}$, it is sufficient to substitute the uniform background field into equation (37), and the lowest mode of localized azimuthal field is found to be 
\begin{equation}
q_{1} \ = \ B_{0}J(\frac{2r_{c}^{2}}{15r^{2}}-\frac{1}{3})
\end{equation}
for $r>r_{c}$, and
\begin{equation}
q_{1} \ = \ B_{0}J(-\frac{8r^{3}}{15r_{c}^{3}}+\frac{r^{4}}{3r_{c}^{4}})
\end{equation}
for $r<r_{c}$. This azimuthal part can work as a source of the localized poloidal field $p_{0}$ in (36), and we obtain 
\begin{equation}
p_{0} \ = \ \frac{B_{0}r_{c}^{3}}{675r}(\frac{r_{c}^{3}}{r^{3}}-\frac{45r_{c}}{4r}+\frac{104}{7})
\end{equation}
for $r>r_{c}$, and
\begin{equation}
p_{0} \ = \ \frac{B_{0}r^{2}}{675}(-\frac{4r^{3}}{r_{c}^{3}}+\frac{45r^{4}}{28r_{c}^{4}}+7)
\end{equation}
for $r<r_{c}$. The generated poloidal part $p_{0}(r)$ has a maximum point in the outer region $r>r_{c}$, and the poloidal field lines along which $(1-x^{2})p_{0}(r)$ is constant show a loop structure on the poloidal plane, which is maintained against the ohmic dissipation. In the slow-rotation limit such a modification of the background uniform field due to the generated poloidal field remains small. Nevertheless, we can expect that a remarkable structure of poloidal field lines as a result of the induced excitation appears in the magnetosphere, if the result presented here is extended to a fast-rotation case of the Kerr geometry.

The role of the generated azimuthal field $B_{T}=3\sin\theta\cos\theta q_{1}(r)/\eta r$ with the strength of the order of $B_{0}J/\eta r$ may be astrophysically more important even in the slow-rotation limit. In the presence of azimuthal magnetic fields, outflows of plasma from the central region may be efficiently drived by the Lorentz force. Further, a Poynting flux is excited if azimuthal magnetic fields exist in the magnatosphere, and in our approximation the Poynting flux vector $P^{A}$ ($A=r,\theta$) is given by
\begin{equation}
P^{A} \ = \ \frac{1}{4\pi}F_{Bt}F^{AB} \ , 
\end{equation}
In the asymptotic region $r\gg r_{c}$ the components $P^{r}$ and $P^{\theta}$ are easily estimated to be 
\begin{equation}
P^{r} \ = \ \frac{B_{0}^{2}J^{2}}{2\pi\eta r^{3}}\sin^{2}\theta\cos^{2}\theta \ , 
\end{equation}
and
\begin{equation}
P^{\theta} \ = \ -\frac{B_{0}^{2}J^{2}}{4\pi\eta r^{4}}\sin\theta\cos\theta
(1+\cos^{2}\theta) \ .
\end{equation}
Interestingly, the Poynting flux propagates outward from the central region and converges toward the rotation axis along the curves given by $(1+\cos^{2}\theta)/r=$ const. Though the generation of azimuthal magnetic fields is also possible through ideal and non-relativistic MHD processes, the dynamical balance between the ohmic dissipation and the induced excitation due to the frame-dragging effect can be an origin of the magnetospheric structure responsible for producing high-energetic phenomena in the polar region. 

\section{THE EFFECT OF CHARGE SEPARATION}

In the case of vanishing $u_{\phi}$ no self-excitation of magnetic fields was found, even if an artificial sharp gradient of the frame-dragging angular velocity $\omega$ was assumed. This is mainly because a sufficiently large $\omega$ is not allowed for any gravity around a central object rotating with a limited angular momentum. We also obtain the upper limit of the angular velocity $\Omega$ of plasma by the requirement $(\Omega-\omega)^{2}<\alpha^{2}/g_{\phi\phi}$ (see eq. [16]). Therefore, even in the cases $\Omega\neq\omega$, the $\omega$-$\Omega$ coupling will remain inefficient for self-excited dynamo, at least within the analytical framework developed here. (As was previously mentioned, the existence of growing modes claimed by N\'{u}\~{n}ez (1997) may be a possible way to self-excited dynamo, if his estimation for the diffusion effect can be justified.) However, the situation may crucially change in the presence of charge separation given by equation (9). For the plasma co-rotating with the angular velocity of the background magnetosphere, the value of $u_{\phi}$ can become large without limit near the light cylinder surface, and a large gradient of $u_{\phi}$ (i. e., a large proper charge density $Q$) will be also allowed to occur there. Then, the azimuthal convection current $Qu_{\phi}$ which appears in equation (10) can play a key role for self-excitation of poloidal flux instead of the frame-dragging effect. 

To study the $Q$-$\Omega$ coupling as a mechanism of self-excited dynamo, let us restrict the following discussion to the case of no gravity and assume the plasma angular velocity to behave as $\Omega=\Omega(R)$. (Hereafter, we use the cylindrical coordinates $R$, $Z$ and $\phi$.) Now the modes $\Psi$ and $F$ satisfying equations (11) and (15) can have the forms
\begin{equation}
\Psi \ = \ \psi(R)\cos(kZ)e^{\mu T} \ , \ \ F \ = \ f(R)\sin(kZ)e^{\mu T} \ , 
\end{equation} 
according to the assumed symmetry with respect to the equatorial plane. Using the Lorentz factor given by $\gamma=1/\sqrt{1-(R\Omega)^{2}}$, the special relativistic versions of equations (11) and (15) reduce to   
\begin{equation}
L\psi-R\Omega^{2}\gamma^{2}\frac{d\psi}{dR} \ = \ -kR\Omega\frac{d\gamma}{dR}f \ , 
\end{equation}
and
\begin{equation}
Lf \ = \ -kR\frac{d\Omega}{dR}\gamma\psi \ , 
\end{equation}
where $L$ is the differential operator defined by 
\begin{equation}
L \ = \ \gamma R\frac{d}{dR}(\frac{1}{\gamma R}\frac{d}{dR})-(k^{2}+\gamma\mu) \ . 
\end{equation}

The plasma angular velocity $\Omega$ may be nearly constant in an inner region (i.e., $\Omega\simeq\Omega_{i}$, which is equal to the angular velocity of the background stationary magnetosphere with a rigid rotation), while it should decrease as $R\rightarrow1/\Omega_{i}$. For mathematical simplicity, we represent such a behavior as a sharp gradient in $\Omega$ at $R=R_{c}<1/\Omega_{i}$. (N\'{u}\~{n}ez (1997) has also studied this case in terms of the $\omega$-$\Omega$ coupling without considering charge separation.) By virtue of the discontinuous gap $\Delta\Omega$ the gradients $d\psi/dR$ and $df/dR$ should also have the discontinuous gaps $\Delta(d\psi/dR)$ and $\Delta(df/dR)$ at $R=R_{c}$. Then, using the equality
\begin{equation}
\frac{R\Omega}{\gamma}d\gamma \ = \ -d(R\Omega+\frac{1}{2}\ln\frac{1-R\Omega}{1+R\Omega}) \ , 
\end{equation}
the integrations of equations (47) and (48) over the narrow region with the sharp gradient in $\Omega$ lead to
\begin{equation}
\Delta(\frac{1}{\gamma}\frac{d\psi}{dR}) \ = \ kf_{c}\Delta(R\Omega+\frac{1}{2}\ln\frac{1-R\Omega}{1+R\Omega}) \ , 
\end{equation}
and
\begin{equation}
\Delta(\frac{1}{\gamma}\frac{df}{dR}) \ = \ -k\psi_{c}\Delta(R\Omega) \ , 
\end{equation}
where $\psi=\psi_{c}$ and $f=f_{c}$ at $R=R_{c}$. 

For further analysis we consider the discontinuous change from $\Omega=\Omega_{i}$ (i.e., $\gamma=1/\sqrt{1-R_{c}^{2}\Omega_{i}^{2}}\equiv\gamma_{i}$) to $\Omega=0$ (i.e., $\gamma=1$). Then, for the modes with $k\gg 1/R_{c}$, the amplitudes of $\psi$ and $f$ should decrease with the forms $\exp(-k_{o}(R-R_{c}))$ at $R>R_{c}$ and $\exp(k_{i}(R-R_{c}))$ at $R<R_{c}$, where 
\begin{equation}
k_{o} \ = \ \sqrt{k^{2}+\mu} \ , 
\end{equation} 
and
\begin{equation}
k_{i} \ = \ \sqrt{k^{2}+\gamma_{i}\mu} \ . 
\end{equation} 
From these behaviors of $\psi$ and $f$ we can easily estimate the discontinuous gaps $\Delta(d\psi/dR)$ and $\Delta(df/dR)$ at $R=R_{c}$, and the growing rate $\mu$ is found to be  
\begin{equation}
\frac{\mu}{k^{2}} \ = \ \frac{a(\gamma_{i})}{b(\gamma_{i})} \ , 
\end{equation}
if the eigenvalue problem of equations (51) and (52) is solved, where
\begin{equation}
a \ = \ \sqrt{\frac{\gamma_{i}-1}{\gamma_{i}+1}}\ln(\gamma_{i}+\sqrt{\gamma_{i}^{2}-1})
-2 \ , 
\end{equation}
and
\begin{equation}
b \ = \ 1+\frac{2}{\gamma_{i}+1}(\gamma_{i}+\frac{(\gamma_{i}-1)^{3/2}}{\sqrt{\gamma_{i}+1}\ln(\gamma_{i}+\sqrt{\gamma^{2}-1})-2\sqrt{\gamma_{i}-1}})^{1/2} \ .
\end{equation}
The stationary mode with $\omega=0$ corresponds to the case $\gamma_{i}=\gamma_{m}$ satisfying $a(\gamma_{m})=0$, and the numerical value is estimated to be $\gamma_{m}\simeq5.55$. 

If plasma rotates with a relativistic velocity giving $\gamma>\gamma_{m}$ in a region around a central object, a sufficient decrease of $\Omega$ within a range $R_{c}(1-\delta)<R<R_{c}(1+\delta)$ ($\delta\ll1$) can excite growing magnetic fields with the scale of $1/k$ much smaller than the distance $R_{c}$. An interesting possibility is that this self-excitation of magnetic fields produces relativistic outflows of plasma across the light cylinder as a result of the back-reaction, though it is a process beyond the scope of the kinematic theory considered in this paper. The spatial variation $\Delta\Omega$ of the angular velocity also may become smaller as the back-reaction works. Then, the self-excited dynamo will stop, and the acceleration of outflows will occur only through a stationary MHD process. The very active phase of self-excitation of magnetic fields and violent plasma acceleration (which will be responsible for flare-like events of radiation flux) can reopen only when a region with highly relativistic angular velocity appears again. 

\section{CONCLUSIONS} 

In a very simplified model we have succeeded in showing the self-excited dynamo through the coupling between $\Omega$ and $Q$ without any frame-dragging effect: Poloidal magnetic field generates azimuthal magnetic field by the help of differential rotation of plasma, and the azimuthal field induces charge separation if the Lorentz factor of the plasma rotation is not constant. The azimuthal convection current carried with the rotating charged plasma can amplify the original poloidal field. Though in our calculation a sharp gradient of $\Omega$ has been assumed, the condition essential to the self-excited dynamo would be the existence of large spatial variations $\Delta(R\Omega)$ and $\Delta\gamma$ such that $\Delta(R\Omega)\Delta\gamma\geq4.5$ within a radial scale $\Delta R$ smaller than $R$. Different from the angular velocity $\omega$ of the dragging of inertial frames, we can expect $\Delta\gamma\gg1$ as an astrophysically permissible case, e.g., if we consider a rotational motion of plasma near the light cylinder in magnetospheres of relativistic compact objects. However, a siginificant poloidal motion may occur near the light cylinder, before $\gamma$ of plasma rotation becomes much larger than unity. Then, the charge separation may be suppressed in the magnetosphere. The self-excited dynamo due to the $\Omega$-$Q$ coupling will be able to work only if strong background fields balance the centrifugal force in a plasma rotating with highly relativistic speed. Hence, this dynamo action should be understood to be an origin of flare-like instability permissible near the light cylinder rather than a mechanism of generation of strong background fields.  

We have also discussed the induced excitation of magnetic fields which occurs through 
the frame-dragging effect acting on uniform background magnetic field. This is a process leading to a new equilibrium configuration of magnetic field lines against the ohmic dissipation. Our analysis has been limited to the case of slow rotation, in which the higher poloidal multipoles remain very small, and the extension to rapid rotation is an important problem to be solved. Though no flare-like activity is expected for this process, the magnetospheric structure with outgoing Pynting flux convergent toward the rotation axis is an interesting result.

In this paper we have clarified only the fundamental aspects of kinematic generation of magnetic fields based on relativistic Ohm's law, and the astrophysical impications for black-hole or neutron-star magnetospheres should be confirmed in more realistic models. In particular, taking account of the presence of poloidal plasma flows will be an important task, which may work for antidynamo. Further, in the cases $\Omega\neq\omega$, the $Q$-$\Omega$ coupling also can be an important origin of the induced excitation even if the self-excited dynamo does not occur. The combined effects of the $\omega$-$\Omega$ and $Q$-$\Omega$ couplings remain unclear in this paper. To treat more appropriately the problem of charge separation, one may need two-component plasma theories (see Khanna 1998a,\markcite{Kh98a} 1998b\markcite{Kh98b}). Axisymmetric dynamo action in charged plasma is an interesting problem in future investigations. 

\acknowledgments

The author thanks Masaaki Takahashi and Masashi Egi for valuable discussions and Ramon Khanna, the referee, for suggestions on improving the manuscript. This work was supported in part by the Grant in-aid for Scientific Research (C) of the Ministry of Education, Science, Sports and Culture of Japan (No.10640257).

\end{document}